H. Sahraoui, H. Mellah, S. Drid, L. Chrifi-Alaoui

# ADAPTIVE MAXIMUM POWER POINT TRACKING USING NEURAL NETWORKS FOR A PHOTOVOLTAIC SYSTEMS ACCORDING GRID

***Introduction.*** *This article deals with the optimization of the energy conversion of a grid-connected photovoltaic system.* ***The novelty*** *is to develop an intelligent maximum power point tracking technique using artificial neural network algorithms.* ***Purpose.*** *Intelligent maximum power point tracking technique is developed in order to improve the photovoltaic system performances under the variations of the temperature and irradiation.* ***Methods.*** *This work is to calculate and follow the maximum power point for a photovoltaic system operating according to the artificial intelligence mechanism is and the latter is used an adaptive modified perturbation and observation maximum power point tracking algorithm based on function sign to generate an specify duty cycle applied to DC-DC converter, where we use the feed forward artificial neural network type trained by Levenberg-Marquardt backpropagation.* ***Results.*** *The photovoltaic system that we chose to simulate and apply this intelligent technique on it is a stand-alone photovoltaic system. According to the results obtained from simulation of the photovoltaic system using adaptive modified perturbation and observation – artificial neural network the efficiency and the quality of the production of energy from photovoltaic is increased.* ***Practical value.*** *The proposed algorithm is validated by a dSPACE DS1104 for different operating conditions. All practice results confirm the effectiveness of our proposed algorithm.* References 37, table 1, figures 27.

***Вступ**. У статті йдеться про оптимізацію перетворення енергії фотоелектричної системи, підключеної до мережі. **Новизна** полягає у розробці методики інтелектуального відстеження точок максимальної потужності з використанням алгоритмів штучної нейронної мережі. **Мета**. Методика інтелектуального відстеження точок максимальної потужності розроблена з метою поліпшення характеристик фотоелектричної системи в умовах зміни температури та опромінення. **Методи**. Робота полягає в обчисленні та відстеженні точки максимальної потужності для фотоелектричної системи, що працює відповідно до механізму штучного інтелекту, і в останній використовується адаптивний модифікований алгоритм збурення та відстеження точок максимальної потужності на основі знаку функції для створення заданого робочого циклу стосовно DC-DC перетворювача, де ми використовуємо штучну нейронну мережу типу «прямої подачі», навчену зворотному розповсюдженню Левенберга-Марквардта. **Результати**. Фотоелектрична система, яку ми обрали для моделювання та застосування цієї інтелектуальної методики, є автономною фотоелектричною системою. Відповідно до результатів, отриманих при моделюванні фотоелектричної системи з використанням адаптивних модифікованих збурень та спостереження – штучної нейронної мережі, ефективність та якість виробництва енергії з фотоелектричної енергії підвищується. **Практична цінність**. Запропонований алгоритм перевірено dSPACE DS1104 для різних умов роботи. Усі практичні результати підтверджують ефективність запропонованого нами алгоритму.* Бібл. 37, табл. 1, рис. 27.
*Ключові слова:* **штучна нейронна мережа, підключена до мережі, адаптивне модифіковане збурення та спостереження, штучна нейронна мережа-відстеження точки максимальної потужності.**

**Introduction.** Nowadays, the electric power generation mainly uses fossil and fissile (nuclear) fuels. The widespread use of fossil fuels, such as gasoline, coal or natural gas, allows for low production prices. On the other hand, their use results in a large release of greenhouse gases and polluting gases. Electricity production from fossil fuels has a great responsibility for global $CO_2$ emissions, hence pollution, according to the last International Energy Agency report [1]. Nuclear power, which does not directly release carbon dioxide, the risks of accident linked to their exploitation are very low but the consequences of an accident would be disastrous. Although the risks of accident linked to their exploitation are very low, but the consequences of an accident would be disastrous and we must not forget the Fukushima Daiichi nuclear disaster in Japan. Furthermore, the treatment of waste from this mode of production is very expensive; the radioactivity of the treated products remains high for many years [2], that's what prompted to the propose a nuclear plant waste management policies and strategies [2], and some researcher suggests to build a regional and global nuclear security system [3]. Finally, uranium reserves are like those of limited oil [4].

Although the world is in surplus in electricity production today, the future is therefore not promising on fossil fuel resources whose reserves are constantly decreasing and whose prices fluctuate enormously depending on the economic situation [5]. The future preparations in the fields of energy production to satisfy the humanity needs should be foreseen today, in order to be able to gradually face the inevitable energy changes.

Each innovation and each breakthrough in research will only have repercussions in about ten years at best, the time to carry out the necessary tests and to consider putting into production without risk for the user as much for his own health than for its electrical installations, to avoid the problems of pollution in the production of electricity, alternative solutions can be photovoltaic (PV), wind, or even hydroelectric sources [4, 5].

The use of PV solar energy seems to be a necessity for the future. Indeed, solar radiation constitutes the most abundant energy resource on earth. The amount of energy released by the Sun, for one hour could be enough to cover global energy needs for a year, for that we should better exploit this energy and optimize its collection by PV collectors [6].

The basic element of a PV system is the solar panel which is made up of photosensitive cells connected to each other. Each cell converts the rays from the Sun into continuous type electricity. PV panels have a specific highly non-linear electrical characteristic which appears clearly in the current-voltage and power-voltage curves [7]. Its electrical characteristics have a particular point called Maximum Power Point (MPP). This point is the optimal operating point for which the panel operates at its maximum power, MPP is highly dependent on climatic conditions and load, which makes the position of the MPP variable over time and therefore difficult to locate [8].

© H. Sahraoui, H. Mellah, S. Drid, L. Chrifi-Alaoui

A Maximum Power Point Tracking (MPPT) control is associated with an intermediate adaptation stage, allowing the PV to operate at the MPP so as to continuously produce the maximum power of PV, whatever the weather conditions (temperature and irradiation), and whatever the charge. The converter control places the system at MPP this point defined by current $I_{mpp}$ and voltage $V_{mpp}$. There are several MPPT techniques that aim to extract maximum power from the solar cells outputs [9-12] the interested reader is referred to [11] for more details. Classic techniques such as the Incremental Conductance technique and Perturbation and Observation (P&O) technique, these two methods are the most used and easy to implement methods but have drawbacks [10-13].

New techniques based on artificial intelligence, such as Fuzzy Logic Control [14, 15] Squirrel Search Algorithm [16], Particle Swarm Optimization [17], Levy Flight Optimization [12], Artificial Neural Networks (ANN) [18, 19], and other propose a hybrid techniques [20-23].

The methods based on ANN allow solving non-linear problems and more complicated by a very fast way since they are represented by non-linear mathematical functions [24]. A different ANN-MPPT algorithm for maximization of power PV production have been studied in many research papers [18, 19, 25]. Messalti et al. in [19] proposes with experimental validation two versions of ANN-MPPT controllers either with fixed or variable step. The aim of their works was to propose an optimal MPPT controller based on neural network for used it in the PV system. Different operating climatic conditions are investigated in the ANN training step in order to improve, tracking accuracy, response time and reduce a chattering.

Kumar et al. in [26] propose two Neural Networks (NN) in the purpose of PV grid-connected with multi-objective and distributed system; one is for assuring MPPT and the other for the generation of reference currents; the NN used for MPPT is based on hill climbing learning algorithm, and use a NN version of a Power Normalized Kernel Least Mean Fourth algorithm control (PNKLMF-NN) to generate a reference currents.

Tavakoliel al. in [27] propose an intelligent method for MPPT control in PV systems, this study establishes a two-level adaptive control framework to increase its efficiency by facilitating system control and efficiently handling uncertainties and perturbations in PV systems and the environment; where the ripple correlation control is the first level of control and the second level is based on an adaptive controller rule for the Model Reference Adaptive Control system and is derived through the use of a self-constructed Lyapunov neural network. However, this approach did not been applied in the purpose of grid connected PV system.

In [28] the authors have made a new technique – a Adaptive Modified Perturbation and Observation (AMPO), which reduces the MPP search steps this last based on the function which widely used in sliding mode control sign function, by this technique of reducing the calculation time and the chattering; compared to classic P&O technique.

Many authors study the issues of PV system grid connection [26], [29-31]. Slama et al. in [29] offer a clever algorithm for determining the best hours to switch between battery and PVs, on the other hand, Belbachir et al. in [30] seeks the optimal integration, both for distributed PVs and for the batteries, other deal with the management of electricity consumption [31].

In this paper, we propose an adaptive P&O algorithm technique based on neural network with the PV system to increase the power of PV and operate at MPP, whatever the climatic variation such as the radiation and the temperature, according a grid demand.

The main contribution of this work is to present an Adaptive Modified Perturbation and Observation – Artificial Neural Networks (AMPO-ANN) based on the sign function which simplifies and reduces the step and time of calculating MPPT point which allows minimizing both the calculation time, the structure of the classic P&O algorithm and the AMPO-ANN designing. Furthermore, in the purpose of real-time application the major problem in a neural network framework is the large size of the created program who add a difficult to implant, especially for the realization of a complex system with small time constant.

We are mainly interested in the development of a control system based on ANN which allows the continuation of the MPP by simulation and experiments, which allows increasing the performance of the neural MPPT chart compared to the other maximization methods. Figure 1 below presents a proposed system of control.

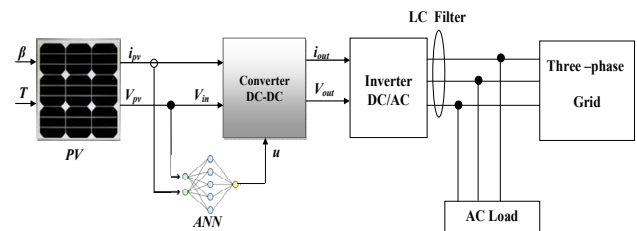

Fig. 1. Schematic diagram of the PV system under study with AMPO-ANN strategy

**The goal of the paper** is to develop a technique of maximization power point tracking search based on the function sign which simplifies and reduces the step size and the computation time of the maximum power point tracking point which minimize both calculation time.

**Subject of investigations.** This paper is valid power maximization technique-based neurons networks by a test bench with a dSPACE DS1104.

**Description and modeling of proposed PV system.** Equation (1) describes the PV cell model, this model (Fig. 2) can be definite by the application of standard data given by the manufacturer. The equivalent circuit for PV cell is presented as follow [32].

The typical equation for a single-diode of PV panel is as follows:

$$I_{pv} = I_{ph} - I_s \left( e^{\frac{V_{pv}+R_s I_{pv}}{aV_T}} - 1 \right) - \frac{V_{pv}+R_s I_{pv}}{R_{sh}}, \quad (1)$$

where $I_{pv}$ is the current generated by PV panel; $I_{ph}$ is the generated photo-current; $I_s$ is the current of saturation; $V_{pv}$ is the voltage of PV panel; $R_s$ is the array's equivalent series resistance; $a$ is the constant of the ideal diode VD; $V_T$ is the thermal voltage of PV ($V_T = KT/q$, where $K$ is the Boltzmann's constant; $T$ is the temperature of PV; $q$ is the charge of an electron); $R_{sh}$ is the array's equivalent parallel resistance.

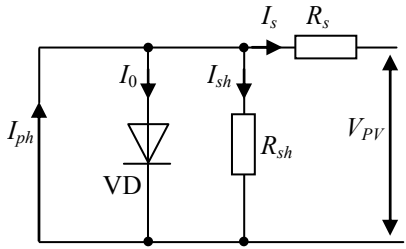

Fig. 2. The PV cell equivalent circuit

Figures 3 and 4 illustrate the PV panel's characteristics as they change; both temperatures between 25 to 75 °C and irradiation between 200 to 1000 W/m² respectively.

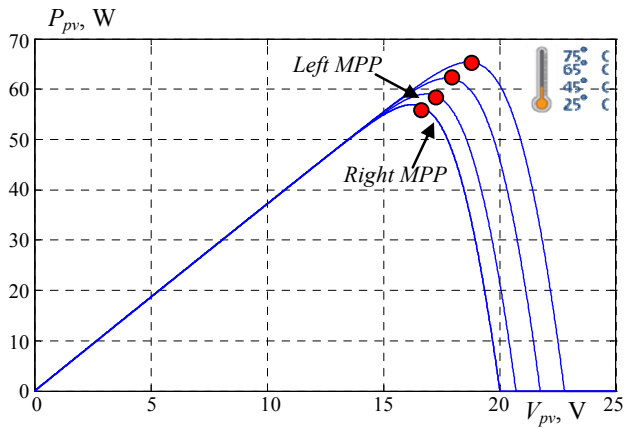

Fig. 3. Characteristics $P = f(V)$ of the PV panel under variation of temperature

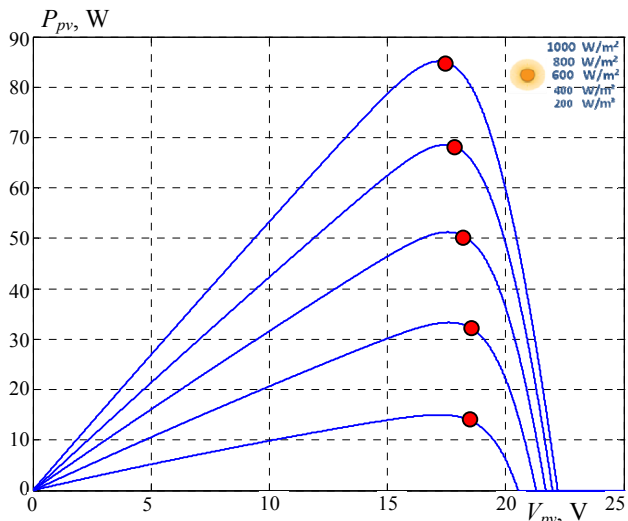

Fig. 4. Characteristics $P = f(V)$ of the PV panel under irradiation variation

**Buck converter modeling.** The load is connected to the DC bus via a DC-DC buck power converter [33] (Fig. 5), which allows it to be controlled.

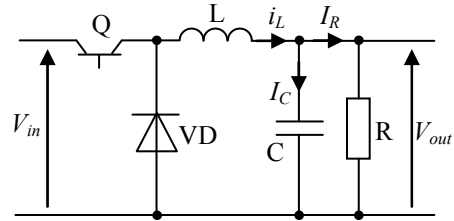

Fig. 5. Buck converter

To model the converter, state space average equations are employed, as shown in the equation (2) [34]

$$\begin{cases} \dot{x}_1 = k_1 u V_{in} - k_1 x_2; \\ \dot{x}_2 = k_2 x_1 - k_3 x_2, \end{cases} \quad (2)$$

where $u$ is duty cycle and $k_1 = 1/L$; $k_2 = 1/C$; $k_3 = 1/RC$.

The steady state is given by
$$[x_1 \quad x_2] = [i_L \quad V_{out}]. \quad (3)$$

• **The DC/AC inverter model.** Figure 6 presents the structure of three-phase voltage source inverter (VSI).

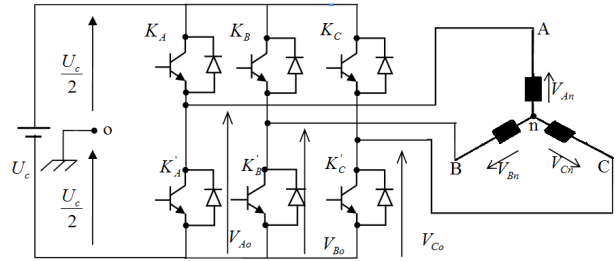

Fig. 6. Structure of a three-phase VSI

The switching function is $C_i\{i = A, B, C\}$ as bellow[35]:
- if $C_i = 1$, then $K_i$ is OFF and $K'_i$ is ON;
- if $C_i = 0$, then $K_i$ is ON and $K'_i$ is OFF.

The outputs voltage of the inverter $U_{AB}$, $U_{BC}$, $U_{CA}$ can be write as:

$$\begin{cases} U_{AB} = U_{Ao} - U_{Bo}; \\ U_{BC} = U_{Bo} - U_{Co}; \\ U_{CA} = U_{Co} - U_{Ao}. \end{cases} \quad (4)$$

Since the phase voltages are star-connected to load sum to zero, equation (4) can be written:

$$\begin{cases} U_{An} = \frac{1}{3}[U_{AB} - U_{CA}]; \\ U_{Bn} = \frac{1}{3}[U_{BC} - U_{AB}]; \\ U_{Cn} = \frac{1}{3}[U_{CA} - U_{BC}]. \end{cases} \quad (5)$$

For the phase-to-neutral voltages of a star-connected load obtain this model:

$$\begin{cases} U_{An} + U_{no} = U_{Ao}; \\ U_{Bn} + U_{no} = U_{Bo}; \\ U_{Cn} + U_{no} = U_{Co}. \end{cases} \quad (6)$$

and we conclude that:

$$U_{no} = \frac{1}{3}(U_{Ao} + U_{Bo} + U_{Co}). \quad (7)$$

For ideal switching can be obtained:
$$U_{io} = C_i \cdot U_c - U_c/2, \qquad (8)$$
with
$$\begin{cases} U_{Ao} = (C_A - 0,5)U_C; \\ U_{Bo} = (C_B - 0,5)U_C; \\ U_{Co} = (C_C - 0,5)U_C. \end{cases} \qquad (9)$$

Substitution of (6) into (7) obtain [30]:
$$\begin{cases} U_{An} = \frac{2}{3}U_{Ao} - \frac{1}{3}U_{Bo} - \frac{1}{3}U_{Co}; \\ U_{Bn} = -\frac{1}{3}U_{Ao} + \frac{2}{3}U_{Bo} - \frac{1}{3}U_{Co}; \\ U_{Cn} = -\frac{1}{3}U_{Ao} - \frac{1}{3}U_{Bo} + \frac{2}{3}U_{Co}. \end{cases} \qquad (10)$$

Setting (9) with (10), obtain:
$$\begin{bmatrix} U_{An} \\ U_{Bn} \\ U_{Cn} \end{bmatrix} = \frac{1}{3} \cdot U_C \begin{bmatrix} 2 & -1 & -1 \\ -1 & 2 & -1 \\ -1 & -1 & 2 \end{bmatrix} \begin{bmatrix} C_A \\ C_B \\ C_C \end{bmatrix}. \qquad (11)$$

**The Conventional P&O Algorithm (CPOA).** In the P&O process the voltage is increased or decreased with a defined step size in the direction of reaching the MPP. The method is carried out again and again till the MPP is attained. In steady condition the operational point oscillates about the MPP, the oscillation is highly dependent on step size, so that when using a small step size, it can reduce volatility but can reduce system dynamics as well. On the other hand, while using a large step size it can improve system dynamics, but it can increase volatility around MPP as well [36]. Figure 7 illustrates the flowchart of the CPOA algorithm.

**AMPO-ANN algorithm.** Many MPPT approaches have recently been created and developed. In terms of accuracy, for real time implementation the P&O MPPT method is more practical than other MPPTs because it is easier to implement [38]. The P&O MPPT technique is primarily based on the perturbation of the PV output voltage $V(t)$ and related output power $P(t)$, which is compared to the prior perturbation $P(t+1)$. Keep the next voltage shift in the same direction as the previous one if the power increases.

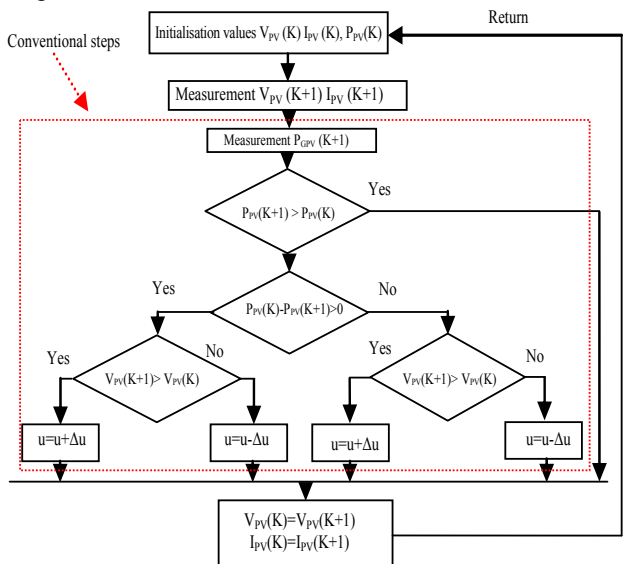

Fig.7. Flowchart of the CPOA algorithm

Artificial intelligence is used in many areas of research, and ANN is a bright and promising part of these technologies, where process control and monitoring, recognition of patterns, power electronics, finance and economics, and medical diagnosis are only a few of the applications where ANNs have proven their worth [37].

In this paper, we will use two neural networks at the same time; the first network whose role is to estimate the output current which corresponds to the maximum power, and the second is used to estimate the voltage which corresponds to the maximum power too [28].

However, if the steps of the algorithm are tracking speed has been increased, as has the accuracy. and rapidity are increased ($dP_{pv}/dV_{pv}>0$), but with high increasing in the oscillation, resulting in comparatively low performance and vice versa, In this paper, an AMPO algorithm method is dedicated to find a simple implantation in comparison with classical CPOA algorithm, and the AMPO can be written as follows:
$$u(\gamma) = U_c(\gamma - 1) + \gamma \cdot \text{sign}(\Delta P), \qquad (12)$$
where $\gamma$ is fixed step and $U_c$ is the voltage control;
$$\Delta P = P(\gamma) - P(\gamma - 1),$$
if $\Delta P > 0$ then increase $U_c$, else $\Delta P < 0$ decrease $U_c$.

A power of the panel ($P_{pv}$) sensor is connected to the P&O algorithm unit in order to detect the power in state $\gamma$ and compare it with next value ($\gamma+1$). At a certain point, when the difference between $P_{pv}(\gamma)$ and $P_{pv}(\gamma+1)$ is $\Delta P_{pv}$ then the algorithm will recognize that there is a powerful change and the algorithm should start from the beginning ($\gamma$). The value of $u(\gamma)$ ($u$ is voltage control of P&O) is set to depend on the value of $\gamma$ of the $\Delta P_{pv}$ criteria and it is different from irradiation values of PV, the $\Delta P_{pv}/\Delta V_{pv}$ change value around at point MPP, the duty cycle follows this change, view the duty cycle varying between values positive, zeros, negative.

In this article, we replace this variation of power $u(\gamma+1)$ show in equation (12) by function $\text{sign}(P_{pv})$ play the role of conventional step of algorithm P&O, with rapidly responses, equation (13) can be written in the following form:
$$\text{sign}(P_{pv}) = \begin{cases} 1 & \text{if} \quad P_{pv} > 0; \\ 0 & \text{if} \quad P_{pv} = 0; \\ -1 & \text{if} \quad P_{pv} < 0; \end{cases} \qquad (13)$$

The adding the variation of power ($\Delta P_{pv}$) and voltage ($\Delta V_{pv}$) can be whiten equation (13) as follow:
$$\text{sign}(\Delta P_{pv}/\Delta V_{pv}) = \begin{cases} 1 & \text{if} \quad \Delta P_{pv}/\Delta V_{pv} > 0; \\ 0 & \text{if} \quad \Delta P_{pv}/\Delta V_{pv} = 0 \text{ at MPP}; \\ -1 & \text{if} \quad \Delta P_{pv}/\Delta V_{pv} < 0. \end{cases} \qquad (14)$$

Equation (14) can be written as follow:
$$\delta = \text{sign}\{(P_{pv}(\gamma) - P_{pv}(\gamma+1)) \cdot (V_{pv}(\gamma) - V_{pv}(\gamma+1))\}. \qquad (15)$$

To simplify the writing of equation (15) can be written in the following form:
$$\delta = \text{sign}(\Delta P_{pv} \cdot \Delta V_{pv}). \qquad (16)$$

State of the voltage control $\delta$ of AMPO can be summarized in Table 1.

Table 1
Variation of MPP in algorithm

| sign($\Delta P_{pv}(\gamma)$) | sign($\Delta P_{pv}(\gamma+1)$) | $\delta$ | $u$ duty cycle | State of MPOA |
|---|---|---|---|---|
| –1 | –1 | –2 | +1 | Left MPP |
| –1 | +1 | 0 | 0 | at MPP |
| +1 | –1 | 0 | 0 | at MPP |
| +1 | +1 | +2 | +1 | Right MPP |

The Table 1 presented the variation of MPP point in algorithm by 4 cases:

- **Case 1**. If state of changing algorithm is sign($\Delta P_{pv}(\gamma)$) = –1, then sign($\Delta P_{pv}(\gamma+1)$) = –1, and $\delta$ = –2, the MPP moving to left; there for $u$ = +1, to increase power of PV ($P_{pv}$).

- **Case 2**. If state of changing algorithm is sign($\Delta P_{pv}(\gamma)$) = +1, then sign($\Delta P_{pv}(\gamma+1)$) = –1, and $\delta$ = 0, at point MPP; there for $u$ = 0, no changing in the power of PV ($P_{pv}$).

- **Case 3**. If state of changing algorithm is sign($\Delta P_{pv}(\gamma)$) = –1, then sign($\Delta P_{pv}(\gamma+1)$) = +1, and $\delta$ = 0, at point MPP; there for $u$ = 0, no changing in the power of PV ($P_{pv}$).

- **Case 4**. If state of changing algorithm is sign($\Delta P_{pv}(\gamma)$) = +1, then sign($\Delta P_{pv}(\gamma+1)$) = +1, and $\delta$ = +2, the MPP moving to right; there for $u$ = +1, to decrease power of PV ($P_{pv}$).

After this case the variation of $\delta$ and $u$ can be thought in AMPO-ANN for desired voltage regulation (for regulate the desire voltage), as shown in Fig. 8.

The value of $\delta$ presented the variation of power of panel $\Delta P_{pv}$ we can add to value of duty cycle $u$ for adjust at point MPP can be written as follow:

$$u(\gamma) = u(\gamma) + \delta \cdot u(\gamma+1). \qquad (16)$$

After equations (15) – (17) can be designing the flow chart of the MPOA algorithm modified shown in Fig. 8 and presented a new step has determined by previous equation.

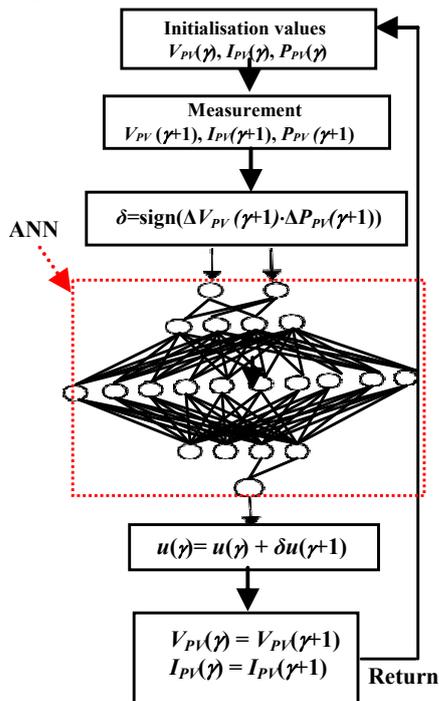

Fig.8. Flowchart of the adaptive ANN-AMPO

**Simulation of proposed system.** The simulation of the Intelligent Maximum Power Point Tracking (IMPPT) based on AMPO-ANN makes it possible to verify that neural networks approach, after learning is effectively capable of predicting the desired output for the values of the data at the input which are not used during learning. We should always compare the true exit from the trajectory of neural networks with the trajectory of the model of PV cells.

The simulations results given in Fig. 9 and represent the electrical characteristics of the stand-alone PV system controlled by AMPO-ANN under standard climatic conditions (1000 W/m² and 25 °C). The powers obtained from the proposed technique stabilize in a steady state around the optimal values delivered by PV ($P_{mpp}$ = 111 W, $V_{mpp}$ = 26 V and $I_{mpp}$ = 4.4 A); AMPO controller allows us for parts per million (PPM) to be attained in 0.06 s, whilst the ANN algorithm allows for PPM to be obtained in 0.02 s only. In addition Fig. 9 shows that in steady state the maximum power supplied by the PV system controlled by the AMPO-ANN is more stable and closer to the PPM compared to AMPO control; the AMPO control give a power oscillates around the MPP which resulting in power losses.

From Fig. 9,*a* we observe that $P_{pv}$ takes 0.02 s in transient state to stabilize at a steady – state value which is MPP in the neighborhood of 111 W.

Figure 9,*b* summarizes a comparison between the MPPT of PV output power controlled by AMPO and AMPO-ANN. We see that the power generalized based on AMPO algorithm has more pikes and is more oscillate compared to the improved technique based on ANN.

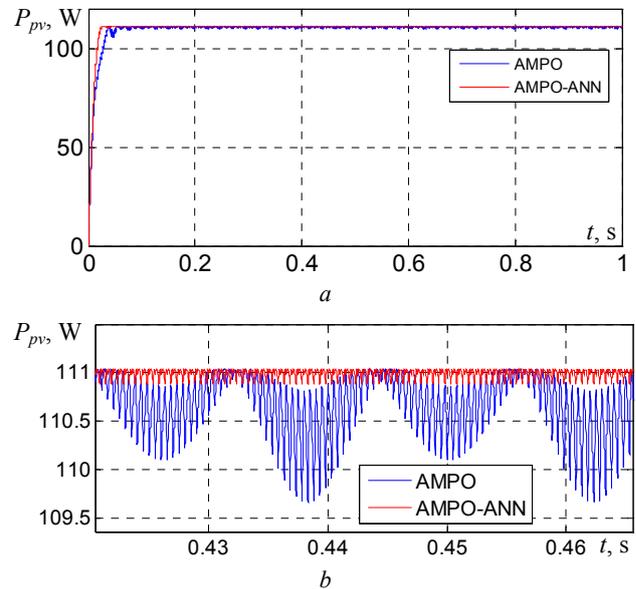

Fig. 9. *a* – power of PV (AMPO-ANN);
*b* – zoom power of PV (AMPO-ANN)

From Fig. 10 we observe the during the period from 0 s to 0.05 s the voltage decreases with significant oscillations, then it stabilizes at the maximum value 26 V.

Figure 11 shows the load current curve based on AMPO-ANN techniques, we note that its value in steady state stabilizes around 4.4 A.

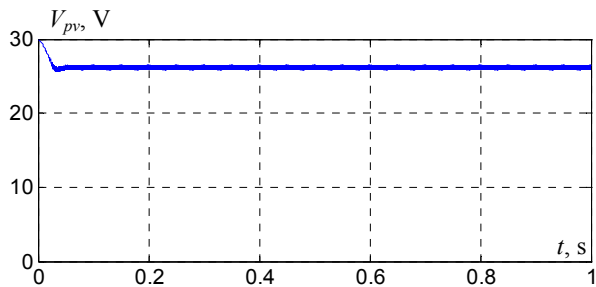
Fig. 10. Voltage of PV (AMPO-ANN)

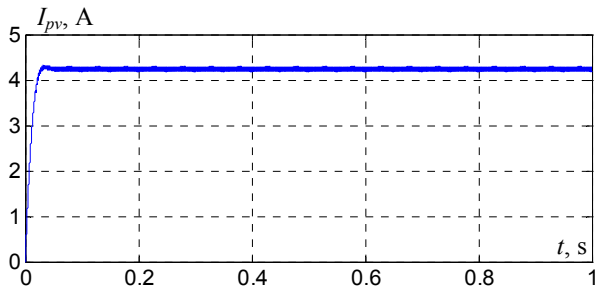
Fig. 11. Current of PV (AMPO-ANN)

In order to verify the robustness and the reliability of the proposed method, we will test the performance of AMPO-ANN by performing separately under climate condition variation, we make variations on solar irradiation and we assume that the temperature is a constant equal to 25 °C, where we suppose that the irradiation drops from 500 to 1000 W/m², at 0.5 s.

According Fig. 12 we note that the maximum power delivered by the PV varies proportionally with irradiation. When the irradiation is 500 W the $P_{pv}$ stabilizes around 38 W. But when the sun goes from 500 to 1000 W/m² the $P_{pv}$ rises to 111 W.

In addition, the simulation result presented by Fig. 9, shows that the AMPO-ANN represent better performances compared to AMPO; since they converge quickly towards the new $P_{mpp}$ with reduced the chattering. Figures 13, 14 show current and voltage of PV based on AMPO-ANN techniques respectively.

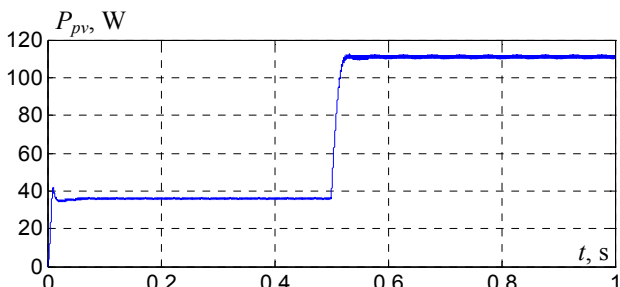
Fig. 12. Power of PV (AMPO-ANN)

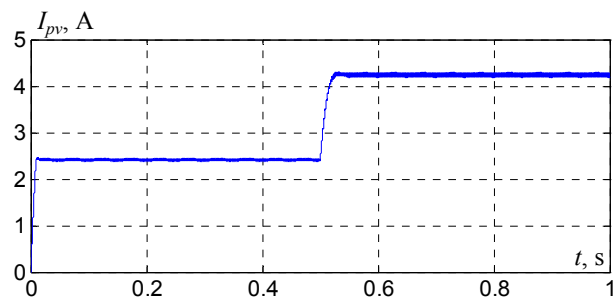
Fig. 13. Current of PV (AMPO-ANN)

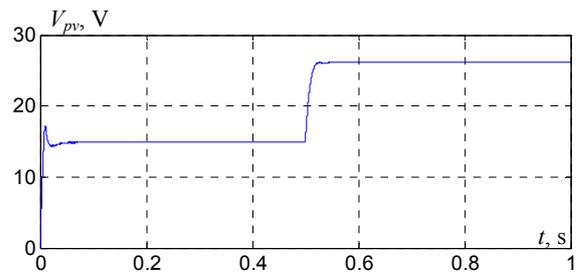
Fig. 14. Voltage of PV (AMPO-ANN)

We will test the performances of the AMPO-ANN algorithm previously developed with the purpose of the grid connection and the climatic conditions are fixed in standard conditions, then connect the PV system to the electrical networks. Figure 15 shows the simulation results. We observe better results for $I_{mes}$ (current measured by the network), $I_{ch}$ (load current) also the three-phase currents $I_a$, $I_b$, $I_c$ (Fig. 16).

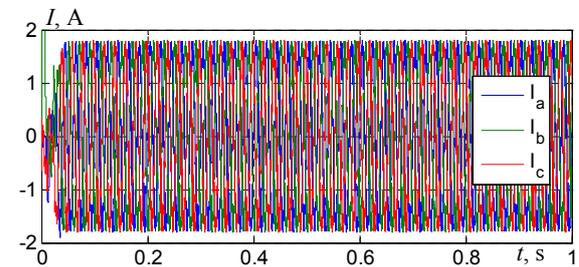

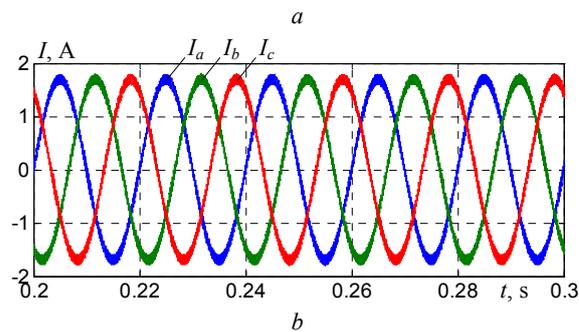
Fig. 15. $a$ – three-phase currents (AMPO-ANN);
$b$ – zoom three-phase currents (AMPO-ANN).

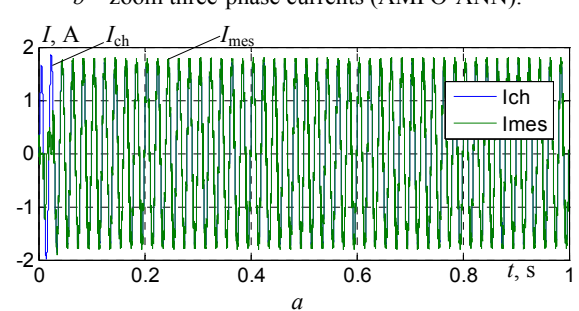

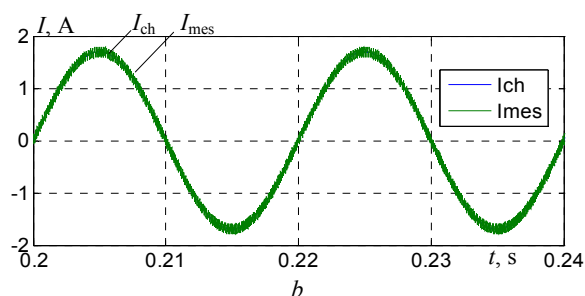
Fig. 16. $a$ – current $I_{ch}$ and $I_{mes}$ (AMPO-ANN);
$b$ – zoom current $I_{ch}$ and $I_{mes}$ (AMPO-ANN)

The results confirm the correct functioning of the two controllers AMPO and AMPO-ANN, but also show a better functioning of the AMPO-ANN. The latter has proven to have better performance, fast response time and very low, steady state error, and it is robust to variations in atmospheric conditions.

**Experimental results.** The proposed AMPO-ANN controller has been put to the test in order to improve its performance. Instead of a solar panel, an experimental setup of a system made of a PV emulator coupled to a DC-DC converter is shown in Fig. 17. LA-25NP and LV-25P are sensors of the current $I_{pv}$ and voltage $V_{pv}$. The proposed control is implemented on the dSPACE DS1104.

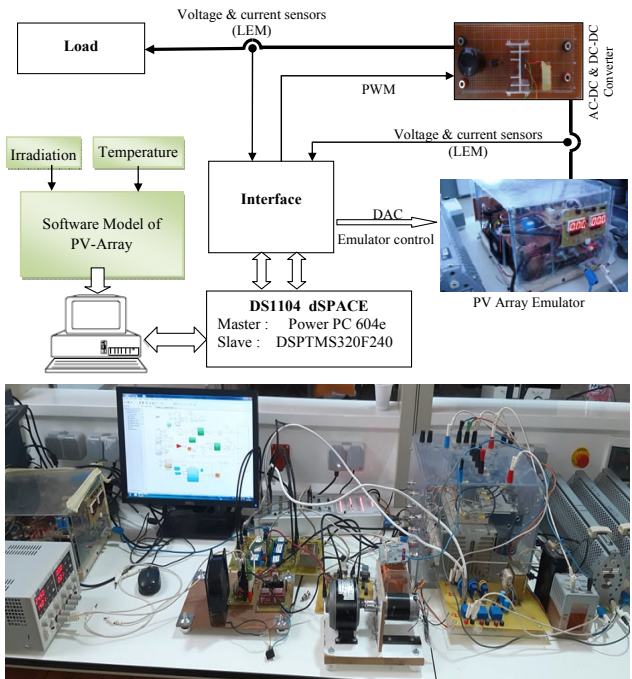

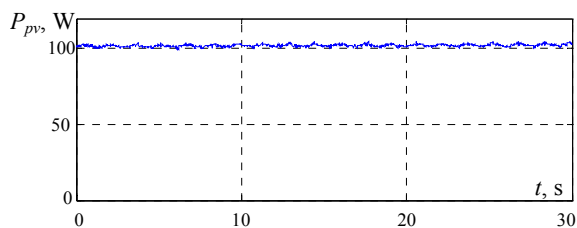
Fig. 17. Structure of the laboratory setup

In the simulation part we assume that all components are perfect (simplifying assumptions, losses and switching phenomena are ignored), so the DC-DC & AC-DC converters has an almost perfect operation.

On the other hand, the tests which we carried out in the laboratory take into account the saturation of the used components and the switching phenomena, these tests consist to validating the proposed technique which applied to a DC-DC converter then connected to an AC-DC converter (inverter).

From Fig. 18-20 we observe that $P_{pv}$ takes 0 s in transient state to stabilize at a steady – state value which is MPP in the neighborhood of 100 W, current and voltage also taken point MPP at values 3.4 A and 29 V.

Fig. 18. Power of PV (AMPO-ANN)

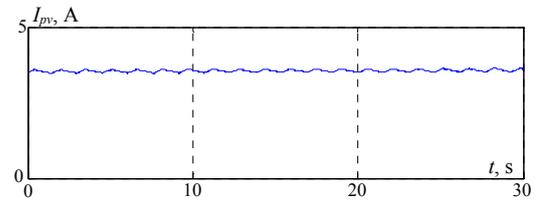
Fig. 19. Current of PV (AMPO-ANN)

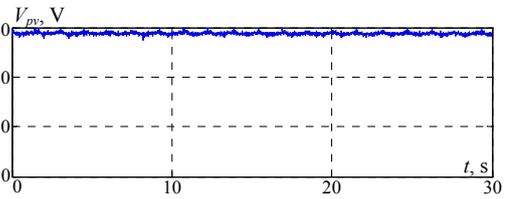
Fig. 20. Voltage of PV (AMPO-ANN)

We will test the performances of the AMPO-ANN algorithm previously developed with the purpose of the grid connection and the climatic conditions are fixed in standard conditions, then connect the PV system to the electrical networks. Figures 21, 22 show the experimental results. We observe better results for $I$ (current measured by the network), $V_{AC}$ (load voltage), also single phase of current and voltage.

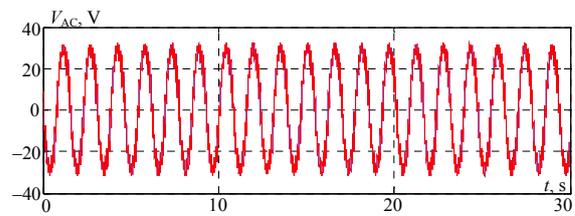
Fig. 21. Voltage of AC bus (AMPO-ANN)

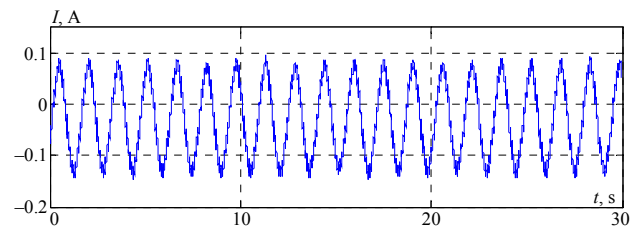
Fig. 22. Current of AC bus (AMPO-ANN)

The Figures 23-27 show the results of realizing the output power of the PV, its operating voltage and current, and the duty cycle (at the frequency of 3000 Hz) for the AMPO-ANN and the conventional disturbance and observation (P&O) AMPO-ANN using a converter and inverter environmental conditions. It is clearly seen how the AMPO-ANN algorithm reduces the response time of the PV system. Obviously, the system with AMPO has a great loss of energy in the transient state, that when the increase in power is the result of the increase in illumination in sinusoidal form between 500 W/m$^2$ and 1000 W/m$^2$, the reversal of the direction of illumination produced by the AMPO-ANN algorithm causes the increase of the power at the MPP point to 101 W and at the same time the output voltage of the inverter, the output voltage $V_{AC}$ is 28 V and current at 0.15 A are illuminate also the harmonics in grid connected, the MPP starts close to the operating point, but the P&O algorithm detects that and moves the operating point in the right direction. The AMPO-ANN algorithm gives a better result than the classic algorithm AMPO.

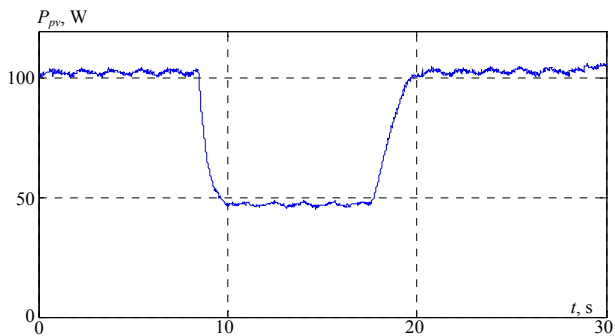
Fig. 23. Power of PV (AMPO-ANN)

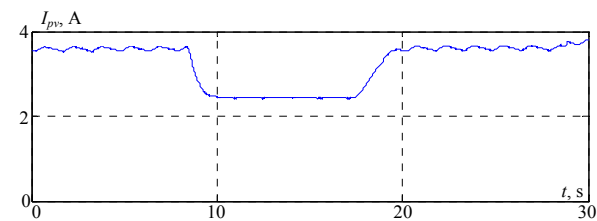
Fig. 24. Current of PV (AMPO-ANN)

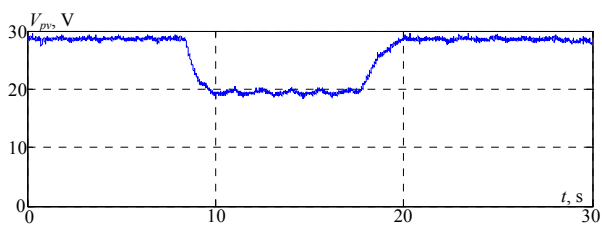
Fig. 25. Voltage of PV (AMPO-ANN)

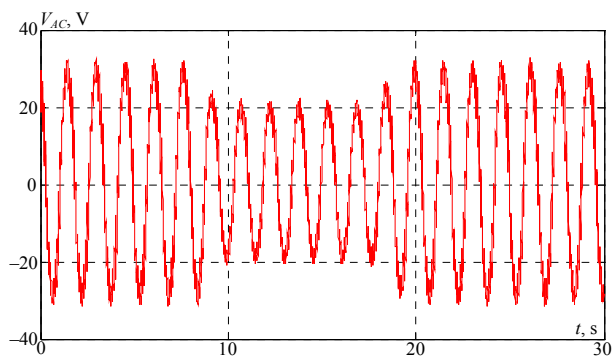
Fig. 26. Voltage of AC bus (AMPO-ANN)

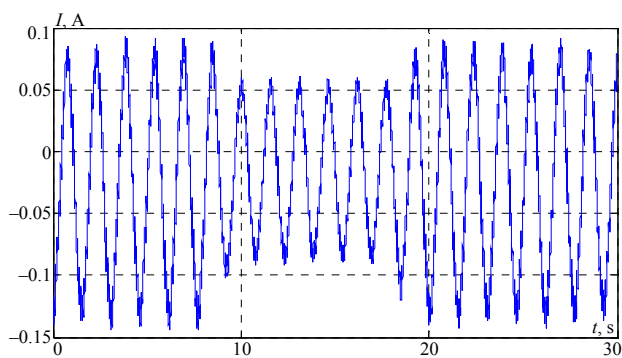
Fig. 27. Current of AC bus (AMPO-ANN)

**Conclusions.**
We analyzed the electrical functioning of a photovoltaic system, adapted by DC-DC converter, regulated by an maximum power point tracking command, to control maximum power point tracking of a photovoltaic system based on neural networks were presented and its architecture of neural networks was used. The simulation and validation results show that this system can adapt the maximum operating point for variations in external disturbances.

We can say that artificial neural networks are efficient and powerful modeling tools, their robustness lies in the possibility of predicting the output of the network even if the relationship with the input is not linear.

The purpose of the modified algorithm adaptive modified perturbation and observation – artificial neural network is to reduce oscillation and achieve a high response of the output power in response to changing weather conditions and parameter variations. All of the results show that the proposed technique control and our improved maximum power point tracking approach are effective.

**Funding.** This work was supported by the Franco-Algerian cooperation program PHC-Maghreb.

**Acknowledgement.** The authors would like to thank laboratory teams of research Propulsion Systems – Electromagnetic Induction, LSPIE, University of Batna 2, Batna, Algeria.

**Conflict of interest.** The authors declare that they have no conflicts of interest.

*Hamza Sahraoui*[1,3], *Doctor of Engineering*,
*Hacene Mellah*[2], *Doctor of Engineering*,
*Said Drid*[3], *Professor, Dr.-Ing. of Engineering*,
*Larbi Chrifi-Alaoui*[4], *Dr.-Ing. of Engineering*,
[1] Electrical Engineering Department,
Hassiba Benbouali University of Chlef,
B.P 78C, Ouled Fares Chlef 02180, Chlef, Algeria,
e-mail: hamzasahraoui@gmail.com
[2] Electrical Engineering Department,
University Akli Mouhand Oulhadj-Bouira,
Rue Drissi Yahia Bouira, 10000, Algeria,
e-mail: has.mel@gmail.com (Corresponding author)
[3] Research Laboratory LSPIE,
Electrical Engineering Department,
University of Batna 2,
53, Route de Constantine, Fésdis, Batna 05078, Algeria,
e-mail: saiddrid@ieee.org
[4] Laboratoire des Technologies Innovantes (LTI),
University of Picardie Jules Verne, IUT de l'Aisne,
13 Avenue François Mitterrand 02880 Cuffies-Soissons, France,
e-mail: larbi.alaoui@u-picardie.fr